\documentclass{scrartcl} 
\usepackage{fixltx2e,fix-cm}
\usepackage{amssymb}
\usepackage{amsmath}
\usepackage{graphicx}
\usepackage{subfigure}
\usepackage{makeidx}
\usepackage{multicol}
\usepackage{lmodern}
\usepackage[a4paper,includeheadfoot,margin=2.54cm]{geometry}

\makeindex

\global\long\def\bt{\bm{\theta}}
\global\long\def\bmu{\bm{\mu}}
\global\long\def\bl{\bm{\lambda}}
\global\long\def\be{\bm{\eta}}
\global\long\def\E{\mathbb{E}}
\global\long\def\by{\bm{y}}
\global\long\def\setY{\mathcal{Y}}
\global\long\def\bS{\bm{\Sigma}}

\global\long\def\ind{\mathbb{I}}
\global\long\def\R{\mathbb{R}}
\global\long\def\bQ{\bm{Q}}
\global\long\def\br{\bm{r}}

\global\long\def\ncore{n_{\mathrm{core}}}

\usepackage[T1]{fontenc}
\usepackage[latin9]{inputenc}
\usepackage{verbatim}
\usepackage{float}
\usepackage{url}
\usepackage{bm}
\usepackage{amsmath}
\usepackage{amssymb}
\usepackage{graphicx}
\usepackage{esint}
\usepackage[authoryear]{natbib}

\makeatletter

\floatstyle{ruled}
\newfloat{algorithm}{tbp}{loa}
\providecommand{\algorithmname}{Algorithm}
\floatname{algorithm}{\protect\algorithmname}

\usepackage{algpseudocode}

\begin{document}

\title{Divide and conquer in ABC: Expectation-Progagation algorithms for likelihood-free inference}
\author{
Simon Barthelm\'e\thanks{CNRS, Gipsa-lab
\texttt{simon.barthelme@gipsa-lab.fr}},
Nicolas Chopin\thanks{
ENSAE-CREST  
\texttt{nicolas.chopin@ensae.fr}} 
Vincent Cottet\thanks{
ENSAE-CREST  
\texttt{vincent.cottet@ensae.fr} }
}

\maketitle

\begin{abstract}
ABC algorithms are notoriously expensive in computing time, as they require simulating many
complete artificial datasets from the model. We advocate in this paper
a ``divide and conquer'' approach to ABC, where we split the likelihood
into $n$ factors, and combine in some way $n$ `local' ABC approximations
of each factor. This has two advantages: (a) such an approach is typically
much faster than standard ABC; and (b) it makes it possible to use
local summary statistics (i.e. summary statistics that depend only
on the data-points that correspond to a single factor), rather than
global summary statistics (that depend on the complete dataset). This
greatly alleviates the bias introduced by summary statistics, and
even removes it entirely in situations where local summary statistics
are simply the identity function. 

We focus on EP (Expectation-Propagation), a convenient and powerful
way to combine $n$ local approximations into a global approximation.
Compared to the EP-ABC approach of \citet{epabc}, we present two variations; 
one based on the parallel EP algorithm of \citet{Cseke2011}, which has
the advantage of being implementable on a parallel architecture; and 
one version which bridges the gap between standard EP and parallel EP. We
illustrate our approach with an expensive application of ABC, namely
inference on spatial extremes.
\end{abstract}

\section{Introduction}

A standard ABC algorithm samples in some way from the pseudo-posterior:
\begin{equation}
p_{\epsilon}^{\mathrm{std}}\left(\bt|\mathbf{y}^{*}\right)\propto p(\bm{\theta})\int p(\mathbf{y}|\bm{\theta})\ind_{\left\{ \left\Vert s(\mathbf{y})-s(\mathbf{\by}^{\star})\right\Vert \leq\epsilon\right\} }\, d\mathbf{y}\label{eq:ABCpost}
\end{equation}
where $p(\mathbf{y}|\bm{\theta})$ denotes the likelihood of data
$\by\in\setY$ given parameter $\bt\in\Theta$, $\mathbf{y}^{*}$
is the actual data, $s$ is some function of the data called a `summary
statistic', and $\epsilon>0$. As discussed elsewhere in this book,
there are various ways to sample from (\ref{eq:ABCpost}), e.g. rejection,
MCMC \citep{Marjoram:MCMCWithoutLikelihood}, SMC \citep{Sisson2007,MR2767283,DelMoral:AdaptSMCforABC},
etc., but they all require simulating a large number of complete datasets
$\by^{j}$ from the likelihood $p(\mathbf{y}|\bm{\theta})$, for different
values of $\bt$. This is typically the bottleneck of the computation.
Another drawback of standard ABC is the dependence on $s$: as $\epsilon\rightarrow0$,
$p_{\epsilon}^{\mathrm{std}}\left(\bt|\mathbf{y}^{*}\right)\rightarrow p(\bt|s(\by^{\star}))\neq p(\bt|\by^{\star})$,
the true posterior distribution, and there is no easy way to choose
$s$ such that $p(\bt|s(\by^{\star}))\approx p(\bt|\by^{\star})$. 

In this paper, we assume that the data may be decomposed into $n$
``chunks'', $\bm{y}=(y_{1},\ldots,y_{n}),$ and that the likelihood may
be factorised accordingly:
\[
p(\mathbf{y}|\bm{\theta})=\prod_{i=1}^{n}f_{i}(y_{i}|\bt)
\]
in such a way that it is possible to sample pseudo-data $y_{i}$
from each factor $f_{i}(y_{i}|\bt)$. The objective is to approximate
the pseudo-posterior: 
\[
p_{\epsilon}(\bt|\by^{\star})\propto p(\bt)\prod_{i=1}^{n}\left\{ \int f_{i}(y_{i}|\bt)\ind_{\left\{ \left\Vert s_{i}(y_{i})-s_{i}(y_{i}^{\star})\right\Vert \leq\epsilon\right\} }\, dy_{i}\right\} 
\]
where $s_{i}$ is a ``local'' summary statistic, which depends only
on $y_{i}$. We expect the bias introduced by the $n$ local summary
statistics $s_{i}$ to be much smaller than the bias introduced by
the global summary statistic $s$. In fact, there are practical cases
where we may take $s_{i}(y_{i})=y_{i}$, removing this bias entirely. 

Note that we do not restrict to models such that the chunks $y_i$ are independent.
In other words, we allow each factor $f_i$ to implicitly depends on other data-points. For
instance, we could have a Markov model, with $f_i(y_i|\bt)=p(y_i|y_{i-1},\bt)$,
or even a model with a more complicated dependence structure, say 
$f_i(y_i|\bt)=p(y_i|y_{1:i-1},\bt)$. The main requirement, however, 
is that we are able to sample from each factor $f_i(y_i|\bt)$. For instance,
in the Markov case, this means we are able to sample from the model realisations of variable $y_i$, conditional on $y_{i-1}=y_{i-1}^\star$ and $\bt$.

Alternatively, in cases where the likelihood does not admit a simple
factorisation, one may replace it by some factorisable pseudo-likelihood;
e.g. a marginal composite likelihood: 
\[
p^{\mathrm{MCL}}(\by|\bt)=\prod_{i=1}^{n}p(y_{i}|\bt)
\]
where $p(y_{i}|\bt)$ is the marginal density of variable $y_{i}$.
Then one would take $f_{i}(y_{i}|\bt)=p(y_{i}|\bt)$ (assuming we are able
to simulate from the marginal distribution
of $y_{i}$). Conditional distributions may be used as well; see \citet{varin2011overview}
for a review of composite likelihoods. Of course, replacing the likelihood
by some factorisable pseudo-likelihood adds an extra level of approximation,
and one must determine in practice whether the computational benefits
are worth the extra cost. Estimation based on composite likelihoods is
generally consistent, but their use in a Bayesian setting results
in posterior distributions that are overconfident (the variance is
too small, as dependent data are effectively treated as independent
observations). 

Many authors have taken advantage of factorisations to speed up ABC.
ABC strategies for hidden Markov models are discussed in \citet{dean2014parameter}
and \citet{yildirim2014parameter}; see the review of \cite{jasra2015approximate}. \citet{White2013} describe a
method based on averages of pseudo-posteriors, which in the Gaussian
case reduces to just doing one pass of parallel EP. \citet{ABCWithCompScoreFunctions}
use composite likelihoods to define low-dimensional summary statistics. 

We focus on Expectation Propagation (EP, \citealp{minka2001expectation}),
a widely successful algorithm for variational inference. In \citet{epabc},
we showed how to adapt EP to a likelihood-free setting. Here we extend
this work with a focus on a parallel variant of EP \citep{Cseke2011}
that enables massive parallelisation of ABC inference. 
For textbook descriptions of EP, see e.g. Section 10.7 of \cite{Bishop:book} or
Section 13.8 of \cite{Gelman:BDA3}. 

The chapter is organised as follows. Section \ref{sec:EP-algorithms}
gives a general presentation of both sequential and parallel EP algorithms.
Section \ref{sec:Applying-EP-ABC} explains how to adapt these EP
algorithms to ABC contexts. It discusses in particular some ways to
speed up EP-ABC. Section \ref{sec:Applications-to-spatial} discusses
how to apply EP-ABC to spatial extreme models. Section \ref{sec:Conclusion}
concludes. 

We use the following notations throughout: bold symbols refer to vectors
or matrices, e.g. $\bt$, $\bl$, $\bS$. For data-points, we use (bold) $\by$ to 
denote complete datasets, and $y_i$ to denote data "chunks", although we
do not necessarily assume the $y_i$'s to be scalars. The letter $p$ typically
refers to probability densities relative to the model: $p(\bt)$ is the prior, 
$p(y_1|\bt)$ is the likelihood of the first data chunk, and so on. The transpose
of matrix $\bf{A}$ is denoted $\mathbf{A}^t$. 

\section{EP algorithms\label{sec:EP-algorithms}}

\subsection{General presentation}

Consider a posterior distribution $\pi(\bt)$ that may be decomposed
into $(n+1)$ factors: 
\[
\pi(\bt)\propto\prod_{i=0}^{n}l_{i}(\bt)
\]
where, say, $l_{0}(\bt)$ is the prior, and $l_{1},\ldots,l_{n}$
are $n$ contributions to the likelihood. Expectation-Progagation
\citep[EP,][]{minka2001expectation} approximates $\pi$ by a similar
decomposition 
\[
q(\bt)\propto\prod_{i=0}^{n}q_{i}(\bt)
\]
where each `site' $q_{i}$ is updated in turn, conditional on the
other factors, in a spirit close to a coordinate-descent algorithm. 

To simplify this rather general framework, one often assumes that
the $q_{i}$ belong to some exponential family of distributions $\mathcal{Q}$
\citep{Seeger:EPExpFam}: 
\[
q_{i}(\bt)=\exp\left\{ \bm{\lambda}_{i}^{t}\bm{t}\left(\bt\right)-\phi\left(\bm{\lambda}_{i}\right)\right\} 
\]
where $\bl_{i}\in\R^{d}$ is the natural parameter, $\bm{t}\left(\bt\right)$
is some function $\Theta\rightarrow\R^{d}$, and $\phi$ is known
variously as the \emph{log-partition function }or the \emph{cumulant
function}: $\phi(\bl)=\log\left[\int\exp\left\{ \bm{\lambda}^{t}\bm{t}\left(\bt\right)\right\} \, d\bt\right]$.
Working with exponential families is convenient for a number of reasons.
In particular, the global approximation $q$ is automatically in the
same family, and with parameter $\bl=\sum_{i=0}^{n}\bl_{i}$: 
\[
q(\bt)\propto\exp\left\{ \left(\sum_{i=0}^{n}\bl_{i}\right)^{t}\bm{t}(\bt)\right\} .
\]

The next section gives additional properties of exponential families
upon which EP relies. Then Section \ref{sub:Site-update} explains
how to perform a site update, that is, how to update $\bl_{i}$, conditional
on the $\bl_{j}$, $j\neq i$, so as, informally, to make $q$ progressively
closer and closer to $\pi$.

\subsection{Properties of exponential families }

Let $\mathrm{KL}(\pi||q)$ be the Kullback-Leibler divergence of $q$
from $\pi$: 
\[
\mathrm{KL}(\pi||q)=\int\pi(\bt)\log\frac{\pi(\bt)}{q(\bt)}\, d\bt.
\]
For a generic member $q_{\bl}(\bt)=\exp\left\{ \bl^{t}\bm{t}(\bt)-\phi(\bl)\right\} $
of our exponential family $\mathcal{Q}$, we have: 
\begin{equation}
\frac{d}{d\bm{\lambda}}\mathrm{KL}(\pi||q_{\lambda})=\frac{d}{d\bm{\lambda}}\phi\left(\bm{\lambda}\right)-\int\pi(\bt)\bm{t}(\bt)\, d\bt\label{eq:minKL}
\end{equation}
 where the derivative of the partition function may be obtained as:
\begin{equation}
\frac{d}{d\bm{\lambda}}\phi\left(\bm{\lambda}\right)=\int\bm{t}(\bt)\exp\left\{ \bm{\lambda}^{t}\bm{t}\left(\bt\right)-\phi(\bl)\right\} \, d\bt=\E_{\bl}\left\{ \bm{t}(\bt)\right\} .\label{eq:moment-params-1}
\end{equation}

Let $\be=\be(\bl)=\E_{\bl}\left\{ \bm{t}(\bt)\right\} $; $\be$ is
called the moment parameter, and there is a one-to-one correspondence
between $\bl$ and $\be$; abusing notations, if $\be=\be(\bl)$ then
$\bl=\bl(\be)$. One may interpret (\ref{eq:minKL}) as follows: finding
the $q_{\bl}$ closest to $\pi$ (in the Kullback-Leibler sense) amounts
to perform \emph{moment matching}, that is, to set $\bl$ such that
the expectation of $\bm{t}(\bt)$ under $\pi$ and under $q_{\bl}$
match. 

To make this discussion more concrete, consider the Gaussian case:
\[
q_{\bl}(\bt)\propto\exp\left\{ -\frac{1}{2}\bt^{t}\bm{Q}\bt+\bm{r}^{t}\bt\right\} ,\quad\bl=\left(\bm{r},-\frac 1 2\bm{Q}\right),
\quad t(\bt)=\left(\bt,\bt\bt^{t}\right)
\]
and the moment parameter
is $\bm{\eta}=\left(\bmu,\bS+\bmu\bmu^t\right)$, 
with $\bS=\bm{Q}^{-1}$, $\bmu=\bm{Q}^{-1}\bm{r}$.
(More precisely, $\bt^t\bQ\bt=\mathrm{trace}(\bQ\bt\bt^t)=\mathrm{vect}(\bQ)^t \mathrm{vect}(\bt\bt^t)$, so the second component of $\bl$ (respectively $\bm{t}(\bt)$)
should be $-{(1 / 2)}\mathrm{vect}(\bQ)$ (resp. $\mathrm{vect}(\bt\bt')$). 
But, for notational convenience, our derivations will be in terms of matrices $\bQ$ and $\bt\bt'$, rather than their vectorised versions.)

In the Gaussian case, minimising $\mathrm{KL}(\pi||q_{\bl})$ amounts to
take $\bl$ such that the corresponding moment parameter  $\left(\bmu,\bS+\bmu\bmu^t\right)$ is such that 
 $\bmu=\E_{\pi}[\bt]$, $\bS=\mathrm{Var}_{\pi}[\bt]$. We will
focus on the Gaussian case in this paper (i.e. EP computes iteratively
a Gaussian approximation of $\pi$), but we go on with the more general
description of EP in terms of exponential families, as this allows
for more compact notations, and also because we believe that other
approximations could be useful in the ABC context.

\subsection{Site update\label{sub:Site-update}}

We now explain how to perform a site update for site $i$, that is,
how to update given $\bl_{i}$, assuming $\left(\bl_{j}\right)_{j\neq i}$
is fixed. Consider the `hybrid' distribution: 
\begin{align*}
h(\bt)\propto q(\bt)\frac{l_{i}(\bt)}{q_{i}(\bt)} & =l_{i}(\bt)\prod_{j\neq i}q_{j}(\bt)\\
 & =l_{i}(\bt)\exp\left\{ \left(\sum_{j\neq i}\bl_{j}\right)^{t}\bm{t}(\bt)\right\} ;
\end{align*}
that is, $h$ is obtained by replacing site $q_{i}$ by the true factor
$l_{i}$ in the global approximation $q$. The hybrid can be viewed as a ``pseudo-posterior'' distribution, formed of the product of a ``pseudo-prior'' $q_{i}$ and a single likelihood site $l_{i}$. The update of site $i$
is performed by minimising $\mathrm{KL}(h||q)$ with respect to $\bl_{i}$
(again, assuming the other $\bl_{j}$, $j\neq i$, are fixed). Informally,
this may be interpreted as a local projection (in the Kullback-Leibler
sense) of $\pi$ to $\mathcal{Q}$. 

Given the properties of exponential families laid out in the previous
section, one sees that this site update amounts to setting $\bl_{i}$
so that $\bl=\sum_{j}\bl_{j}$ matches $\mathbb{E}_{h}[\bm{t}(\bt)]$,
the expectation of $t(\bt)$ with respect to the hybrid distribution.
In addition, one may express the update of $\bl_{i}$ as a function
of the current values of $\bl_{i}$ and $\bl$, using the fact that 
$\sum_{j\neq i} \bl_j=\bl-\bl_i$, 
as done below in Algorithm \ref{alg:Generic-site-update}.

\begin{algorithm}[h]
\protect\caption{\label{alg:Generic-site-update}Generic site update in EP}

Function $\mathrm{SiteUpdate}(i,l_{i},\bl_i,\bl)$:

1. Compute 
\[
\bl^{\mathrm{new}}:=\bl\left(\mathbb{E}_{h}[\bm{t}(\bt)]\right),
\quad\bl_{i}^{\mathrm{new}}:=\bl^{\mathrm{new}}-\bl+\bl_i
\]
where $ \be \rightarrow \bl(\be)$ is the function that maps the moment parameters
to the natural parameters (for the considered exponential family, see previous section) and 
\begin{equation}
\mathbb{E}_{h}\left[\bm{t}(\bt)\right]=
\frac{\int \bm{t}(\bt) l_{i}(\bt)\exp\left\{ \left(\bl-\bl_i\right)^{t}\bm{t}(\bt)\right\} \, d\bt}
{\int l_{i}(\bt)\exp\left\{ \left(\bl-\bl_i\right)^{t}\bm{t}(\bt)\right\} \, d\bt}
.\label{eq:EP_moment_matching}
\end{equation}

2. Return $\bl_{i}^{\mathrm{new}}$, and optionally $\bl^{\mathrm{new}}$
(as determined by syntax, i.e. either $\bl_{i}^{\mathrm{new}}\leftarrow\mathrm{SiteUpdate}(i,l_{i},\bl_i,\bl)$,
or $\left(\bl_{i}^{\mathrm{new}},\bl^{\mathrm{new}}\right)\leftarrow\mathrm{SiteUpdate}(i,l_{i},\bl_i,\bl)$).
\end{algorithm}

In practice, the feasibility of EP for a given posterior is essentially
determined by the difficulty to evaluate, or approximate, the integral
(\ref{eq:EP_moment_matching}). Note the simple interpretation of
this quantity: this is the posterior expectation of $\bm{t}(\bt)$,
for pseudo-prior $q_{-i}$, and pseudo-likelihood the likelihood factor
$l_{i}(\bt)$. (In the EP literature, the pseudo-prior $q_{-i}$ is often called
the cavity distribution, and the pseudo-posterior $\propto q_{-i}(\bt)l_i(\bt)$
the tilted or hybrid distribution.)

\subsection{Gaussian sites}

In this paper, we will focus on Gaussian approximations; that is $\mathcal{Q}$
is the set of Gaussian densities
\[
q_{\bl}(\bt)\propto\exp\left\{ -\frac{1}{2}\bt^{t}\bm{Q}\bt+\bm{r}^{t}\bt\right\} ,\quad\bl=\left(\bm{r},-\frac 1 2 \bm{Q}\right)
\]
and EP computes iteratively a Gaussian approximation of $\pi$, obtained
as a product of Gaussian factors. For this particular family, simple
calculations show that the site updates take the form given by Algorithm
\ref{alg:EP-Site-update-Gaussian-case}. 

\begin{algorithm}[h]
\protect\caption{EP Site update (Gaussian case)\label{alg:EP-Site-update-Gaussian-case}}

Function $\mathrm{SiteUpdate}(i,l_{i},\left(\bm{r}_{i},\bQ_{i}\right),\left(\bm{r},\bQ\right))$:
\begin{enumerate}
\item Compute 
\begin{align*}
Z_{h} & =\int q_{-i}(\bt)l_{i}(\bt)\, d\bt\\
\bmu_{h} & =\frac{1}{Z_{h}}\int\bt q_{-i}(\bt)l_{i}(\bt)\, d\bt\\
\bS_{h} & =\frac{1}{Z_{h}}\int\bt\bt^{t}q_{-i}(\bt)l_{i}(\bt)\, d\bt-\bmu_{h}\bmu_{h}^{t}
\end{align*}
where $q_{-i}(\bt)$ is the Gaussian density
\[
q_{-i}(\bt)\propto \exp\left\{ -\frac{1}{2}\bt^{t}\left(\bQ-\bQ_i \right)\bt+\left(\br-\br_i\right)^{t}\bt\right\} .
\]

\item Return $\left(\br_{i}^{\mathrm{new}},\bQ_{i}^{\mathrm{new}}\right)$,
and optionally $\left(\br^{\mathrm{new}},\bQ^{\mathrm{new}}\right)$
(according to syntax as in Algorithm \ref{alg:Generic-site-update}),
where 
\begin{align*}
\left(\bQ^{\mathrm{new}},\br^{\mathrm{new}}\right) & =\left(\bS_{h}^{-1},\bS_{h}^{-1}\bmu_{h}\right),\\
\left(\bQ_{i}^{\mathrm{new}},\br_{i}^{\mathrm{new}}\right) & =\left(\bQ_{i}+\bQ^{\mathrm{new}}-\bm{Q},\br_{i}+\br^{\mathrm{new}}-\bm{\br}\right).
\end{align*}
 \end{enumerate}
\end{algorithm}

In words, one must compute the expectation and variance of the pseudo-posterior
obtained by multiplying the Gaussian pseudo-prior $q_{-i}$, and likelihood
$l_{i}$.

\subsection{Order of site updates: sequential EP, parallel EP, and block-parallel
EP\label{sub:Order-of-site}}

We now discuss in which \emph{order }the site updates may be performed;
i.e. should site updates be performed sequentially, or in parallel,
or something in between. 

The initial version of EP, as described in \citet{minka2001expectation},
was purely sequential (and will therefore be referred to as ``sequential
EP'' from now on): one updates $\bl_{0}$ given the current values
of $\bl_{1},\ldots,\bl_{n}$, then one updates $\bl_{1}$ given $\bl_{0}$
(as modified in the previous update) and $\bl_{2},\ldots,\bl_{n}$,
and so on; see Algorithm \ref{alg:Sequential-EP}. Since the function
$\mathrm{SiteUpdate}\left(i,l_{i},\bl_{i},\bl\right)$ computes the
updated version of both $\bl_{i}$ and $\bl=\sum_{j=0}^{n}\bl_{j}$,
$\bl$ changes at each call of $\mathrm{SiteUpdate}$. 

\begin{algorithm}[h]
\protect\caption{\label{alg:Sequential-EP}Sequential EP}

\begin{algorithmic}
\Require initial values for $\bl_0,\ldots,\bl_n$
\State $\bl \leftarrow \sum_{i=0}^n \bl_i$ 
\Repeat
\For{$i=0$ to $n$}
\State $\left( \bl_i, \bl \right) \leftarrow
\mathrm{SiteUpdate}\left(i,l_i,\bl_i,\bl\right)$
\EndFor
\Until convergence
\State \Return $\bl$
\end{algorithmic}
\end{algorithm}

Algorithm \ref{alg:Sequential-EP} is typically run until $\bl=\sum_{i=0}^{n}\bl_{i}$
stabilises in some sense. 

The main drawback of sequential EP is that, given its sequential nature,
it is not easily amenable to parallel computation. \citet{Cseke2011}
proposed a parallel EP algorithm, where all sites are updated in parallel,
independently of each other. This is equivalent to update the sum
$\bl=\sum_{i=0}^{n}\bl_{i}$ only after all the sites have been updated;
see Algorithm \ref{alg:parallel-EP}. 

\begin{algorithm}[h]
\protect\caption{\label{alg:parallel-EP}Parallel EP}

\begin{algorithmic}
\Require initial values for $\bl_0,\ldots,\bl_n$
\State $\bl \leftarrow \sum_{i=0}^n \bl_i$ 
\Repeat
\For{$i=0$ to $n$} (parallel) 
\State $\bl_i \leftarrow
\mathrm{SiteUpdate}\left(i,l_i,\bl_i,\bl \right)$ 
\EndFor
\State $\bl\leftarrow \sum_{i=0}^n \bl_i$ 
\Until convergence 
\State \Return $\bl$
\end{algorithmic}
\end{algorithm}

Parallel EP is ``embarrassingly parallel'', since its inner loop
performs $(n+1)$ independent operations. A drawback of parallel EP
is that its convergence is typically slower (i.e. requires more complete
passes over all the sites) than sequential EP. Indeed, during the
first pass, all the sites are provided with the same initial global
approximation $\bl$, whereas in sequential EP, the first site updates
allow to refine progressively $\bl$, which makes the following updates
easier. 

We now propose a simple hybrid of these two EP algorithms, which we
call block-parallel EP. We assume we have $\ncore$
cores (single processing units) at our disposal. 
For each block of $\ncore$ successive
sites, we update these $\ncore$ sites in parallel, and then update
the global approximation $\bl$ after these $\ncore$ updates;
see Algorithm \ref{alg:block-parallel-EP}. 

\begin{algorithm}
\protect\caption{\label{alg:block-parallel-EP}Block-parallel EP}

\begin{algorithmic}
\Require initial values for $\bl_0,\ldots,\bl_n$
\State $\bl \leftarrow \sum_{i=0}^n \bl_i$ 
\Repeat
\For{$k=1$ to $\lceil (n+1)/\ncore\rceil$}
\For{$i=(k-1)\ncore$ to $(k\ncore-1)\wedge n$} (parallel)
\State $\bl_i \leftarrow \mathrm{SiteUpdate}\left(i,l_i,\bl_i,\bl\right)$
\EndFor
\State $\bl\leftarrow \sum_{i=0}^n \bl_i$ 
\EndFor
\Until convergence 
\State \Return $\bl$
\end{algorithmic}
\end{algorithm}

Quite clearly, block-parallel EP generalises both sequential EP (take
$\ncore=1$) and parallel EP (take $\ncore=n+1$). This generalisation
is useful in any situation where the actual number of cores $\ncore$
available in a given architecture is such that $\ncore\ll(n+1)$.
In this way, we achieve essentially the same speed-up as Parallel
EP in terms of parallelisation (since only $\ncore$ cores are available
anyway), but we also progress faster thanks to the sequential nature
of the successive block updates. We shall discuss more specifically
in the next section the advantage of block-parallel EP over standard
parallel EP in an ABC context.

\subsection{Other practical considerations}

Often, the prior, which was identified with $l_{0}$ in our factorisation,
already belongs to the approximating parametric family: $p(\bt)=q_{\bl_{0}}(\bt)$.
In that case, one may fix beforehand $q_{0}(\bt)=l_{0}(\bt)=p(\bt)$,
and update only $\bl_{1},\ldots,\bl_{n}$ in the course of the algorithm,
while keeping $\bl_{0}$ fixed to the value given by the prior. 

EP also provides at no extra cost an approximation of the normalising
constant of $\pi$: $Z=\int_{\bt}\prod_{i=0}^{n}l_{i}(\bt)\, d\bt$.
When $\pi$ is a posterior, this can be used to approximate the marginal
likelihood (evidence) of the model. See e.g. \citet{epabc} for more
details. 

In certain cases, EP updates are ``too fast'', in the sense that
the update of difficult sites may lead to e.g. degenerate precision
matrices (in the Gaussian case). One well known method to slow down
EP is to perform fractional updates \citep{minka2004power}; that is, informally, update only
a fraction $\alpha\in(0,1]$ of the site parameters; see Algorithm
\ref{alg:site-update-alpha}. 

\begin{algorithm}[h]
\protect\caption{\label{alg:site-update-alpha}Generic site update in EP (fractional
version, requires $\alpha\in(0,1]$)}

Function $\mathrm{SiteUpdate}(i,l_{i},\bl_{i},\bl)$: 

1. Compute 
\[
\bl^{\mathrm{new}}:=\alpha\bl\left(\mathbb{E}_{h}[\bm{t}(\bt)]\right)+(1-\alpha)\bl,\quad\bl_{i}^{\mathrm{new}}:=\bl_{i}+\alpha\left\{ \bl\left(\mathbb{E}_{h}[\bm{t}(\bt)]\right)-\bl\right\} 
\]
 with $\mathbb{E}_{h}[\bm{t}(\bt)]$ defined in \eqref{eq:moment-params-1},
 see Step 1 of \ref{alg:Generic-site-update}.

2. As Step 2 of Algorithm \ref{alg:Generic-site-update}. 
\end{algorithm}

{} 

In practice, reducing $\alpha$ is often the first thing to try when
EP either diverges or fails because of non-invertible matrices (in
the Gaussian case). Of course, the price to pay is that with a lower
$\alpha$, EP may require more iterations to converge.

\subsection{Theoretical properties of EP}

EP is known to work well in practice, sometimes surprisingly so, but it has proved quite resistant to theoretical study. In \citet{epabc} we could give no guarantees whatsoever, but since then the situation has improved. 
The most important question concerns the quality of the approximations produced by EP. Under relatively strong conditions \citet{DehaeneBarthelme:EPLargeDataLimit} were able to show that Gaussian EP is asymptotically exact in the large-data limit. This means that if the posterior tends to a Gaussian (which usually happens in identifiable models), then EP will recover the exact posterior. \cite{DehaeneBarthelme:BoundingErrorsEP} show further that EP recovers the mean of the posterior with an error that vanishes in $\mathcal{O}(n^{-2})$, where $n$ is the number of data-points. The error is up to an order of magnitude lower than what one can expect from the canonical Gaussian approximation, which uses the mode of the posterior as an approximation to the mean. 

However, in order to have an EP approximation, one needs to find one in the first place. The various flavours of EP (including the ones described here) are all relatively complex fixed-point iterations and their convergence is hard to study. \citet{DehaeneBarthelme:EPLargeDataLimit} show that parallel EP converges in the large-data limit to a Newton iteration, and inherits the potential instabilities in Newton's method. Just like Newton's method, non-convergence in EP can be fixed by slowing down the iterations, as described above. 

The general picture is that EP should work very well if the hybrids are well-behaved (log-concave, roughly). Like any Gaussian approximation it can be arbitrarily poor when used on multi-modal posterior distributions, unless the modes are all equivalent. 

Note finally that the results above apply to variants of EP where hybrid distributions are tractable (meaning their moments can be computed exactly). In ABC applications that is not the case, and we will incur additional Monte Carlo error. As we will explain, part of the trick in using EP in ABC settings is finding ways of minimising that additional source of errors.


\section{Applying EP in ABC\label{sec:Applying-EP-ABC}}

\subsection{Principle}

Recall that our objective is to approximate the ABC posterior 
\[
p_{\epsilon}(\bt|\by^{\star})\propto p(\bt)\prod_{i=1}^{n}\left\{ \int f_{i}(y_{i}|\bt)\ind_{\left\{ \left\Vert s_{i}(y_{i})-s_{i}(y_{i}^{_{\star}})\right\Vert \leq\epsilon\right\} }\, dy_{i}\right\} 
\]
for a certain factorisation of the likelihood, and for a certain collection
of local summary statistics $s_{i}$. This immediately suggests using
EP on the following collection of sites 
\[
l_{i}(\bt)=\int f_{i}(y_{i}|\bt)\ind_{\left\{ \left\Vert s_{i}(y_{i})-s_{i}(y_{i}^{_{\star}})\right\Vert \leq\epsilon\right\} }\, dy_{i}
\]
for $i=1,\ldots,n$. For convenience, we focus on the Gaussian case
(i.e. the $l_{i}$'s will be approximated by Gaussian factors $q_{i}$),
and assume that the prior $p(\bt)$ itself is already Gaussian, and
does not need to be approximated. 

From Algorithm \ref{alg:EP-Site-update-Gaussian-case}, we see that,
in this Gaussian case, it is possible to perform a site update provided
that we are able to compute the mean and variance of a pseudo-posterior,
corresponding to a Gaussian prior $q_{-i}$, and likelihood $l_{i}$. 

Algorithm \ref{alg:Local-ABC-exercise} describes a simple rejection
algorithm that may be used to perform the site update. Using this
particular algorithm inside sequential EP leads to the EP-ABC algorithm
derived in \citet{epabc}. We stress however that one may generally
use any ABC approach to perform such a site update. The main point
is that this local ABC problem is much simpler than ABC 
for the complete likelihood for two reasons. First, the pseudo-prior
$q_{-i}$ is typically much more informative than the true prior $p(\bt)$,
because $q_{-i}$ approximates the posterior of all the data minus
$y_{i}$. Thus, we are much less likely to sample values of $\bt$
with low likelihood. Second, even for a fixed $\bt$, the probability
that $\left\Vert s_{i}(y_{i})-s_{i}(y_{i}^{\star})\right\Vert \leq\epsilon$
is typically much larger than $\left\Vert s(\by)-s(\by^{\star})\right\Vert \leq\epsilon$,
as $s_{i}$ is generally of lower dimension than $s$. 

\begin{algorithm}
\protect\caption{Local ABC algorithm to perform site update\label{alg:Local-ABC-exercise}}

Function $\mathrm{SiteUpdate}(i,f_{i},\left(\bm{r}_{i},\bm{Q}_{i}\right),\left(\bm{r},\bm{Q}\right))$:
\begin{enumerate}
\item Simulate $\bt^{(1)},\ldots,\bt^{(M)}\sim N(\bmu_{-i},\bS_{-i})$ where
$\bS_{-i}^{-1}=\bm{Q}-\bm{Q}_{i}$, $\bmu_{-i}=\bS_{-i}\left(\br-\br_{i}\right)$.
\item For each $m=1,\ldots M$, simulate $y_{i}^{(m)}\sim f_{i}(\cdot|\bt^{(m)}).$
\item Compute
\begin{align*}
M_{\mathrm{acc}} & =\sum_{m=1}^{M}\ind\left\{ \left\Vert s_{i}(y_{i}^{(m)})-s_{i}(y_{i}^{\star})\right\Vert \leq\epsilon\right\} \\
\hat{\bmu}_{h} & =\frac{1}{M_{\mathrm{acc}}}\sum_{m=1}^{M}\bt^{(m)}\ind\left\{ \left\Vert s_{i}(y_{i}^{(m)})-s_{i}(y_{i}^{\star})\right\Vert \leq\epsilon\right\} \\
\hat{\bS}_{h} & =\frac{1}{M_{\mathrm{acc}}}\sum_{m=1}^{M}\bt^{(m)}\left[\bt^{(m)}\right]^{t}\ind\left\{ \left\Vert s_{i}(y_{i}^{(m)})-s_{i}(y_{i}^{\star})\right\Vert \leq\epsilon\right\} -\hat{\bmu}_h\hat{\bmu}_h^{t}
\end{align*}

\item Return $\left(\br_{i}^{\mathrm{new}},\bQ_{i}^{\mathrm{new}}\right)$,
and optionally $\left(\br^{\mathrm{new}},\bQ^{\mathrm{new}}\right)$
(according to syntax as in Algorithm \ref{alg:Generic-site-update}),
where 
\begin{align*}
\left(\bQ^{\mathrm{new}},\br^{\mathrm{new}}\right) & =\left(\hat{\bS}_{h}^{-1},\hat{\bS}_{h}^{-1}\hat{\bmu}_{h}\right),\\
\left(\bQ_{i}^{\mathrm{new}},\br_{i}^{\mathrm{new}}\right) & =\left(\bQ_{i}+\bQ^{\mathrm{new}}-\bm{Q},\br_{i}+\br^{\mathrm{new}}-\bm{\br}\right).
\end{align*}
\end{enumerate}
\end{algorithm}

\subsection{Practical considerations\label{sub:Practical-considerations}}

We have observed that in many problems the acceptance rate of Algorithm
\ref{alg:Local-ABC-exercise} may vary significantly across sites,
so, instead of fixing $M$, the number of simulated pairs $(\bt^{(m)},y_{i}^{(m)})$, 
to a given value, we recommend to sample until the number of accepted
pairs (i.e. the number of $(\bt^{(m)},y_{i}^{(m)})$ such that $\left\Vert s_{i}(y_{i}^{(m)})-s_{i}(y_{i}^{(m)})\right\Vert \leq\epsilon$)
equals a certain threshold $M_{0}$. 

Another simple way to improve EP-ABC is to generate the $\bt^{(m)}$
using quasi-Monte Carlo: for distribution $N(\bmu_{-i},\bS_{-i})$,
we take $\bt^{m}=\bmu_{-i}+\bm{L}\bm{\Phi}^{-1}(\bm{u}^{m})$, where
$\bm{\Phi}^{-1}$ is the Rosenblatt transformation (multivariate quantile function)
of the unit normal distribution of dimension $\mathrm{dim}(\bt)$,
 $\bm{L}\bm{L^{t}}=\bS_{-i}$
is the Cholesky decomposition of $\bS_{-i}$, and the $\bm{u}^{(m)}$
is a low-discrepancy sequence, such as the Halton sequence; see e.g.
Chap. 5 in \citet{Lemieux:MCandQMCSampling} for more background on
low-discrepancy sequences and quasi-Monte Carlo. 

Regarding $\epsilon$, our practical experience is that finding a reasonable value through trial and error is typically much easier with
EP-ABC than with standard ABC. This is because the $y_i$'s are typically of much lower dimension than the complete data-set $\by$. However, one more elaborate recipe to calibrate $\epsilon$ is to run EP-ABC with a first value of $\epsilon$, then set $\epsilon$ to the minimal value such that the proportion of simulated $y_i$ at each site
such that $\|s_i(y_i)-s_i(y_i^\star)\|\leq \epsilon$ is above,
say, $5\%$. Then one may start over with this new value of $\epsilon$. 

Another direction suggested by Mark Beaumont in a personal communication is to correct the estimated precision matrices for bias, using formula (4) from \citet{PazSancher:ImprovingPrecMatric}. 

\subsection{Speeding up parallel EP-ABC in the IID case}

This section considers the IID case, i.e. the model assumes that the
$y_{i}$ are IID (independent and identically distributed), given
$\bt$: then 
\[
p(\by|\bt)=\prod_{i=1}^{n}f_{1}(y_{i}|\bt)
\]
where $f_{1}$ denotes the common density of the $y_{i}$. In this
particular case, each of the $n$ local ABC posteriors, as described
by Algorithm \ref{alg:Local-ABC-exercise}, will use pseudo-data
from the \emph{same }distribution (given $\bt$). This suggests recycling
these simulations across sites. 

\citet{epabc} proposed a recycling strategy based on sequential importance
sampling. Here, we present an even simpler scheme that may be implemented
when Parallel EP is used. At the start of iteration $t$ of Parallel
EP, we sample $\bt^{(1)},\ldots,\bt^{(M)}\sim N(\bmu,\bS)$, the current
\emph{global} approximation of the posterior. For each $\bt^{m}$,
we sample $y^{(m)}\sim f_{1}(y|\bt^{m})$. Then, for each site $i$,
we can compute the first two moments of the hybrid distribution by
simply doing an importance sampling step, from $N(\bmu,\bS)$ to $N(\bmu_{-i},\bS_{-i})$,
which is obtained by dividing the density of $N(\bmu,\bS)$ by factor
$q_{i}$. 
Specifically, the weight function is: 
$$
\frac{|\bQ_{-i}|\exp\left\{ -\frac{1}{2}\bt^{t}\bQ_{-i}\bt+\br_{-i}^{t}\bt\right\} }{|\bQ|\exp\left\{ -\frac{1}{2}\bt^{t}\bQ\bt+\br^{t}\bt\right\} }
=\frac{|\bQ-\bQ_i|}{|\bQ|}\exp\left\{ \frac{1}{2}\bt^{t}\bQ_{i}\bt-\br_{i}^{r}\bt\right\} 
$$
since $\bQ=\bQ_{i}+\bQ_{-i}$, $\br=\br_{i}+\br_{-i}$. Note that
further savings can be obtained by retaining the samples for several
iterations, regenerating only when the global approximation has changed
too much relative to the values used for sampling. In our implementation
we monitor the drift by computing the Effective Sample Size of importance
sampling from $N(\bmu,\bS)$ (the distribution of the current samples)
for the new global approximation $N(\bmu',\bS')$.

We summarise the so-obtained algorithm as Algorithm \ref{alg:Parallel-EP-ABC-recycling}.
Clearly, recycling allows us for a massive speed-up when the number
$n$ of sites is large, as we re-use the same set of simulated pairs
$(\bt^{(m)},y^{(m)})$ for all the $n$ sites. In turns, this allows
us to take a larger value for $M$, the number of simulations, which
leads to more stable results. 

We have advocated Parallel EP in Section \ref{sec:EP-algorithms}
as a way to parallelise the computations over the $n$ sites. Given
the particular structure of Algorithm \ref{alg:Parallel-EP-ABC-recycling},
we see that it is also easy to parallelise the simulation of the $M$
pairs $(\bt^{(m)},y^{(m)})$ that is performed at the start of each
EP iteration; this part is usually the bottleneck of the computation.
In fact, we also observe that Algorithm \ref{alg:Parallel-EP-ABC-recycling}
performs slightly better than the recycling version of EP-ABC (as
described in \citealt{epabc}) even on a non-parallel architecture.

\begin{algorithm}
\begin{algorithmic} 
\Require $M$ (number of samples), initial values for $(\br_i,\bQ_i)_{i=0,\ldots,n}$ (note $(\br_0,\bQ_0)$ stays constant during the course of the algorithm, as we have assumed a Gaussian prior with natural parameter $(\br_0,\bQ_0)$)
\Repeat
\State $\bQ\leftarrow\sum_{i=0}^n \bQ_i$, $\br\leftarrow\sum_{i=0}^n \br_i$, $\bS\leftarrow\bQ^{-1}$, $\bmu\leftarrow\bS\br$
\For{$m=1,\ldots,M$}
\State $\bt^{(m)}\sim N(\bmu,\bS)$
\State $y^{(m)}\sim f_1(y|\bt^{(m)})$ 
\EndFor
\For{$i=1,\ldots,n$}
\For{$m=1,\ldots,M$}
\State $w^{(m)} \leftarrow \frac{|\bQ-\bQ_i|}{|\bQ|} \exp\left\{ \frac 1 2 (\bt^{(m)})^t \bQ_i \bt^{(m)} - \br_i^t \bt^{(m)} \right\}
\ind\left\{ \left\Vert s_{i}(y_{i}^{(m)})-s_{i}(y_{i}^\star)\right\Vert \leq\epsilon\right\}$
\EndFor
\State $\hat{Z}\leftarrow M^{-1}\sum_{m=1}^M w^{(m)}$
\State $\hat{\bmu}\leftarrow (M\hat{Z})^{-1}\times\sum_{m=1}^{M}w^{(m)}\bt^{(m)}$
\State $\hat{\bS}\leftarrow (M\hat{Z})^{-1}\times\sum_{m=1}^{M}w^{(m)}\bt^{(m)}\left[\bt^{(m)}\right]^{t} -\hat{\bmu}\hat{\bmu}^{t}$
\State $\bm{r}_{i}\leftarrow\hat{\bS}^{-1}\hat{\bmu}-\bm{r}_{-i}$
\State $\bm{Q}_{i}\leftarrow\hat{\bS}^{-1}-\bm{Q}_{-i}$ 
\EndFor
\Until{Stopping rule (e.g. changes in $(\br,\bQ)$ have become small)}
\end{algorithmic}\protect\caption{\label{alg:Parallel-EP-ABC-recycling}Parallel EP-ABC with recycling
(IID case)}

\end{algorithm}

\section{Application to spatial extremes\label{sec:Applications-to-spatial}}

We now turn our attention to likelihood-free inference for spatial
extremes, following \citet{MR2892354}, see also \citet{Prangle2014}.

\subsection{Background}

The data $\by$ consist of $n$ IID observations $y_{i}$, typically
observed over time, where $y_{i}\in\mathbb{R}^{d}$ represents some
maximal measure (e.g. rainfall) collected at $d$ locations $x_{j}$
(e.g. in $\mathbb{R}^{2}$). The standard modelling approach for extremes
is to assign to $y_{i}$ a max-stable distribution (i.e. a distribution
stable by maximisation, in the same way that Gaussians are stable
by addition). In the spatial case, the vector $y_{i}$ is composed
of $d$ observations of a max-stable process $x\rightarrow Y(x)$
at the $d$ locations $x_{j}$. A general approach to defining max-stable
processes is \citep{MR1947786}:
\begin{equation}
Y(x)=\max_{k}\left\{ s_{k}\max\left(0,Z_{k}(x)\right)\right\} \label{eq:charac_spatial_proc}
\end{equation}
where $(s_{k})_{k=1}^{\infty}$ is the realisation of a Poisson process
over $\mathbb{R}^{+}$ with intensity $\Lambda(\mathrm{d}s)=\mu^{-1}s^{-2}\mathrm{d}s$
(if we view the Poisson process as producing a random set of ``spikes''
on the positive real line, then $s_{1}$ is the location of the first
spike, $s_{2}$ the second, etc.), $(Z_{k})_{k=1}^{\infty}$ is a
countable collection of IID realisations of a zero-mean, unit-variance
stationary Gaussian process, with correlation function $\rho(h)=\mathrm{Corr}(Z_{k}(x),Z_{k}(x'))$
for $x$, $x'$ such that $\|x-x'\|=h$, and $\mu=\mathbb{E}\left[\max\left(0,Z_{k}(x)\right)\right]$.
Note that $Y(x)$ is marginally distributed according to a unit Fr\'{e}chet
distribution, with CDF $F(y)=\exp(-1/y)$. 

As in \citet{MR2892354}, we will consider the following parametric
Whittle-Mat\'{e}rn correlation function 
\[
\rho_{\bt}(h)=\frac{2^{1-\nu}}{\Gamma(\nu)}\left(\frac{h}{c}\right)^{\nu}K_{\nu}(\frac{h}{c}),\quad c,\nu>0
\]
where $K_{\nu}$ is the modified Bessel function of the third kind.
We take $\bt=(\log\nu,\log c)$ so that $\bm{\Theta}=\mathbb{R}^{2}$.
(We will return to this logarithmic parametrisation later.) 

The main issue with spatial extremes is that, unless $d\leq2$, the
likelihood $p(\by|\bt)$ is intractable. One approach to estimate
$\bt$ is pairwise marginal composite likelihood \citep{MR2757202}.
Alternatively, (\ref{eq:charac_spatial_proc}) suggests a simple way
to simulate from $p(\by|\bt)$, at least approximately (e.g. by truncating
the domain of the Poisson process to $[0,S_{\max}]$). This motivates
likelihood-free inference \citep{MR2892354}.

\subsection{Summary statistics}

One issue however with likelihood-free inference for this class of
models is the choice of summary statistics: \citet{MR2892354} compare
several choices, and find that the one that performs best is some
summary of the clustering of the $d(d-1)(d-2)/6$ triplet-wise coefficients
\[
\frac{n}{\sum_{i=1}^{n}\left\{ \max(y_{i}(x_{j}),y_{i}(x_{k}),y_{i}(x_{l}))\right\} ^{-1}},\quad1\leq j<k<l\leq d.
\]

But computing these coefficients require $\mathcal{O}(d^{3})$ operations,
and may actually be more expensive than simulating the data itself: \citet{Prangle2014}
observes in a particular experiment than the cost of computing these
coefficients is already more that twice the cost of simulating data
for $d=20$. As a result, the overall approach of \citet{MR2892354}
may take several days to run on a single-core computer.

In contrast, EP-ABC allows us to define local summary statistics,
$s_{i}(y_{i})$, that depend only on one data-point $y_{i}$. We simply
take $s_{i}(y_{i})$ to be the (2-dimensional) OLS (ordinary least
squares) estimate of regression 
\[
\log\left|F(y_{i}(x_{j}))-F(y_{i}(x_{k}))\right|=a+b\log\left\Vert x_{j}-x_{k}\right\Vert +\epsilon_{jk},\quad1\leq j<k\leq d
\]
where $F$ is the unit Fr\'{e}chet CDF. The madogram function $h\rightarrow\mathbb{E}\left[\left|Y(x)-Y(x')\right|\right]$,
for $\|x-x'\|=h$, or its empirical version, is a common summary of
spatial dependencies (for extremes). Here, we take the $F-$madogram, i.e. $Y(x)$
is replaced by $F(Y(x))\sim U[0,1]$, because $Y(x)$ is Fr\'{e}chet and
thus $\mathbb{E}\left[\left|Y(x)\right|\right]=+\infty$.

\subsection{Numerical results on real data}

\begin{figure}
\centering{}\includegraphics[scale=0.5]{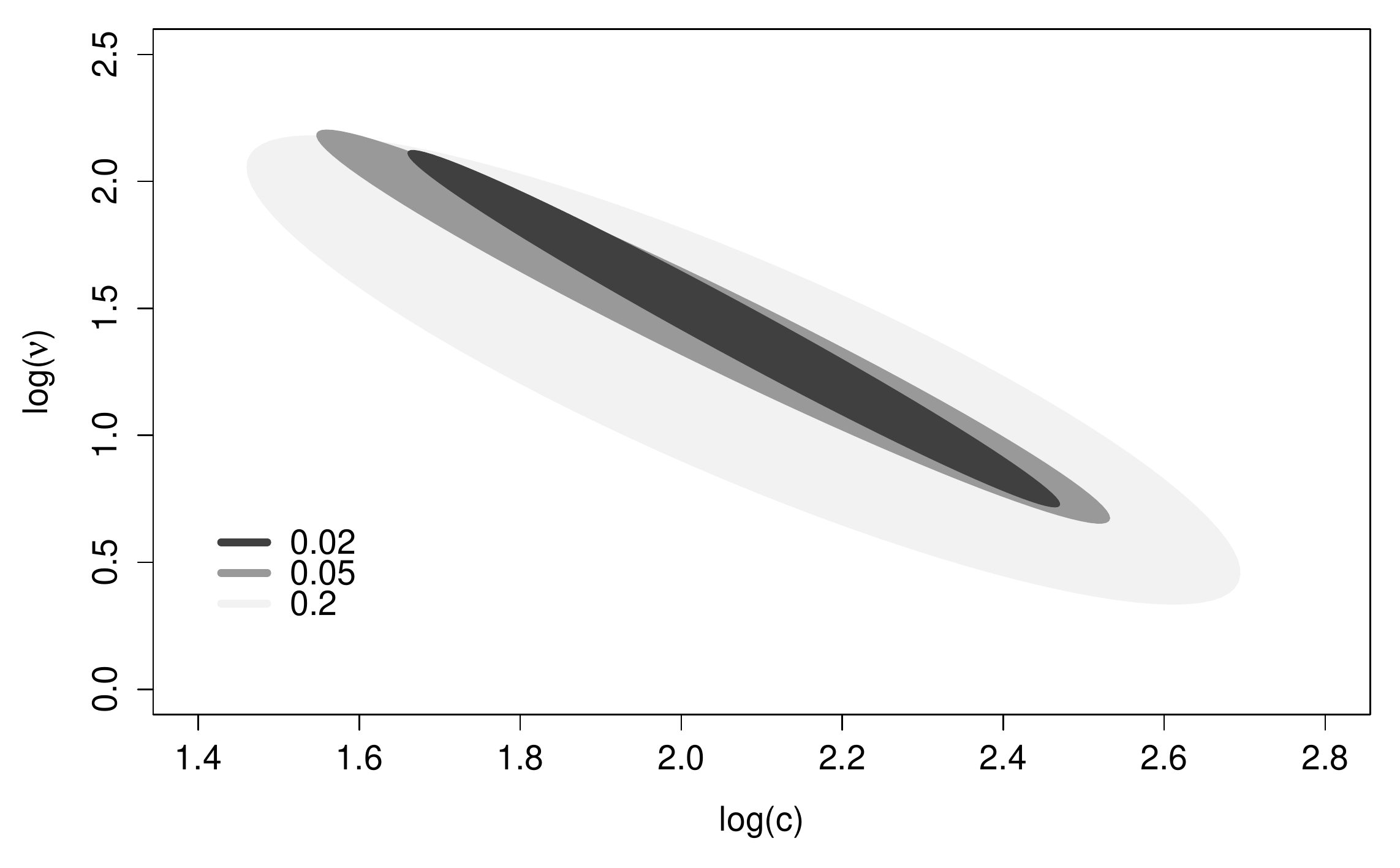}\protect\caption{\label{fig:ellipses}50\% credible ellipses of the Gaussian approximation of the posterior computed by EP-ABC,
for different values of $\epsilon$, and rainfall dataset. }
\end{figure}

We now apply EP-ABC to the rainfall dataset of the \textsf{SpatialExtremes
R} package (available at \url{http://spatialextremes.r-forge.r-project.org/}),
which records maximum daily rainfall amounts over the years 1962\textendash 2008
occurring during June\textendash August at 79 sites in Switzerland.
We ran sequential EP with recycling and quasi-Monte Carlo (see discussion in Section \ref{sub:Practical-considerations}). 
Figure \ref{fig:ellipses} plots the EP-ABC posterior for 
$\epsilon=0.2$, $0.05$ and $0.02$. A $N(0,1)$ prior was used for both components
of $\bt=\left(\log\nu,\log c\right)$. 

Each run took about 3 hours on our desktop computer, and generated
about $10^{5}$ data-points (i.e. realisations $y_i\in\R^d$, where $d$ is the number
of stations). As a point of comparison, we ran \citet{MR2892354}'s \textsf{R} package
for a week on the same computer, which led to the generation of 
$5\times 10^4$ complete datasets (i.e. $\approx 4\times 10^6$ data-points). 
However, the ABC posterior approximation obtained from 
the 100 generated datasets that were closest to the data, relative to their summary statistics,
was not significantly different from the prior.

\begin{figure}
\centering{}\includegraphics[bb=100bp 250bp 500bp 600bp,clip,scale=0.4]{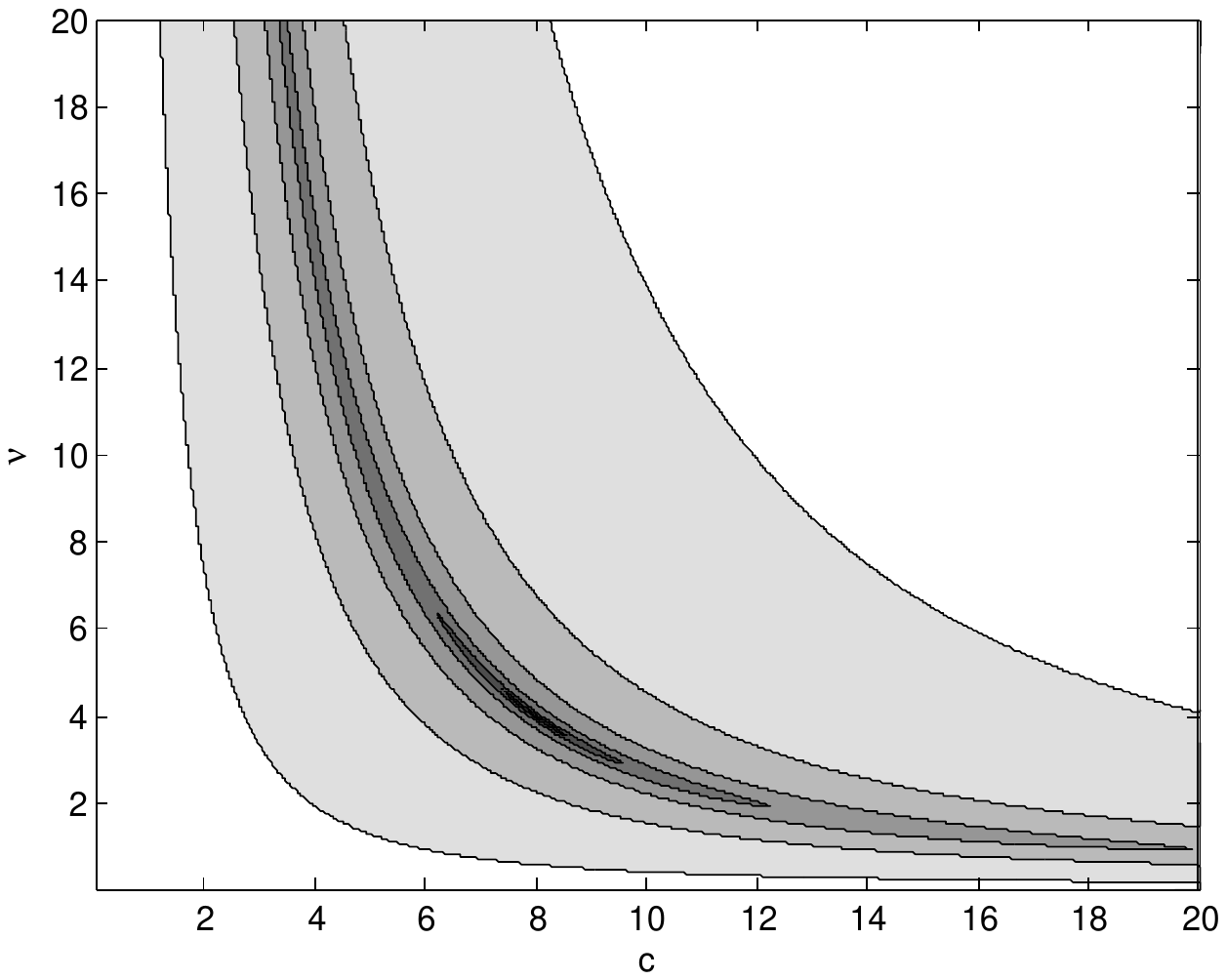}\includegraphics[bb=100bp 250bp 500bp 600bp,clip,scale=0.4]{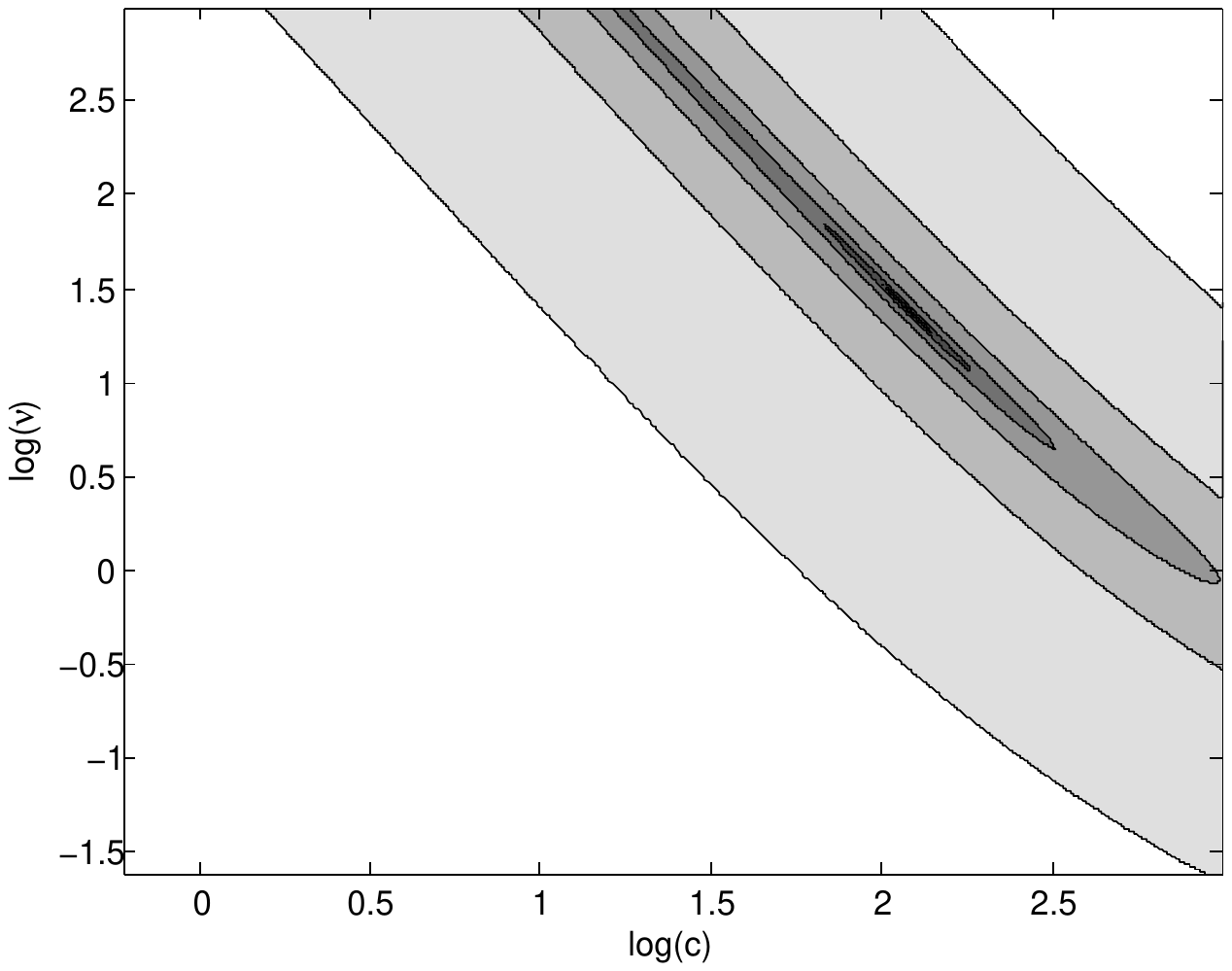}\protect\caption{\label{fig:Heat-map}Heat map of functions $\left(\nu,c\right)\rightarrow\int|\rho_{\nu,c}-\rho_{\nu_{0},c_{0}}|$
and $\left(\log\nu,\log c\right)\rightarrow\int|\rho_{\nu,c}-\rho_{\nu_{0},c_{0}}|$,
for $\left(\nu_{0},c_{0}\right)=\left(8,4\right)$. }
\end{figure}

Finally, we discuss the strong posterior correlations between the
two parameters that are apparent in Figure \ref{fig:ellipses}. Figure
\ref{fig:Heat-map} plots a heat map of functions $\left(\nu,c\right)\rightarrow\int|\rho_{\nu,c}-\rho_{\nu_{0},c_{0}}|$
and $\left(\log\nu,\log c\right)\rightarrow\int|\rho_{\nu,c}-\rho_{\nu_{0},c_{0}}|$,
for $\left(\nu_{0},c_{0}\right)=\left(8,4\right)$. The model appears
to be nearly non-identifiable, as values of $\left(\nu,c\right)$
that are far away may produce correlation functions that are nearly
indistinguishable. In addition, the parametrisation $\bt=\left(\log\nu,\log c\right)$
has the advantage of giving an approximately Gaussian shape to contours,
which is clearly helpful in our case given that EP-ABC generates a
Gaussian approximation. Still, it is interesting to note that EP-ABC
performs well on such a nearly non-identifiable problem. 

\subsection{EP Convergence}

\begin{figure}
\begin{centering}
\includegraphics[bb=100bp 250bp 500bp 600bp,scale=0.4]{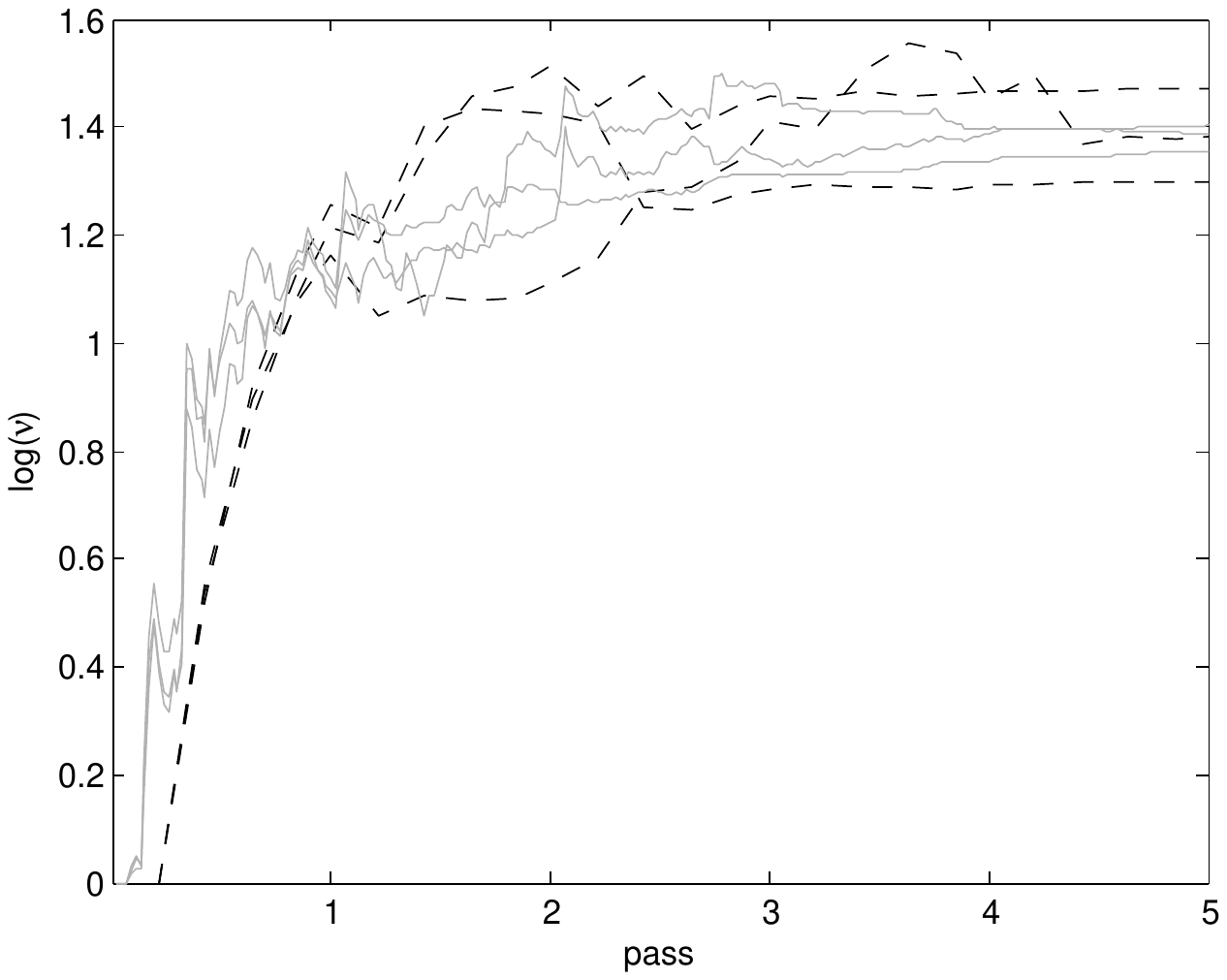}\includegraphics[bb=100bp 250bp 500bp 600bp,scale=0.4]{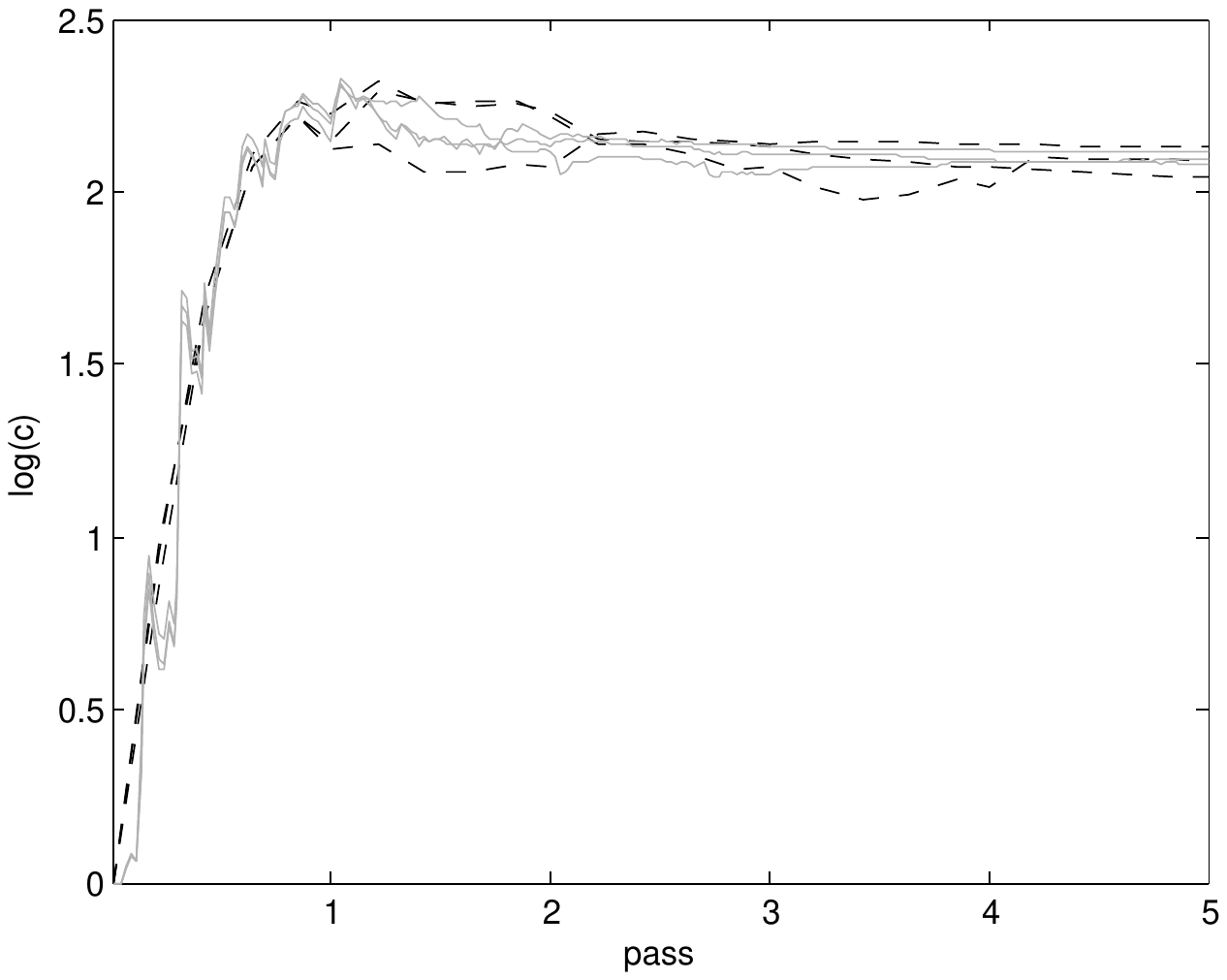}
\par\end{centering}

\protect\caption{\label{fig:conv_EP}Posterior mean of $\log\nu$ (left panel) and $\log c$
(right panel) as a function of the number of passes (one pass equals $n=47$ site updates), 
for 3 runs of the sequential version (solid grey line), and block-parallel version
($n_{\mathrm{core}}=10$, dashed black line) of EP-ABC, applied to rainfall
dataset ($\epsilon=0.05$). }

\end{figure}

Finally, we compare the convergence (relative to the number of iterations)
of the standard version, and the block-parallel version (described
in Section \ref{sub:Order-of-site}) of EP-ABC, on the rainfall dataset
discussed above. Figure \ref{fig:conv_EP} plots the evolution of
the posterior mean of both parameters $\nu$ (left panel) and $c$
(right panel), relative to the number of site updates, for 3 runs
of both versions, and for $\epsilon=0.05$. 

We took $n_{\mathrm{core}}=10$ (i.e. blocks of 10 sites are updated
in parallel), although both algorithms were run on a single core.
We see that both algorithms essentially converge at the same rate.
Thus, if implemented on a 10-core machine, the block-parallel version
should offer essentially a $\times10$ speed-up.

\section{Conclusion\label{sec:Conclusion}}

Compared to standard ABC, the main drawback of EP-ABC is that it introduces
an extra level of approximation, because of its EP component. On the
other hand, EP-ABC strongly reduces, or sometimes removes entirely,
the bias introduced by summary statistics, as it makes possible to
use $n$ local summaries, instead of just one for the complete dataset.
In our experience (see e.g. the examples in \citet{epabc}), this
bias reduction more than compensates the bias introduced by EP. But
the main advantage of EP-ABC is that it is much faster than standard
ABC. Speed-ups of more than $100$ are common, as evidenced by our
spatial extremes example. 

We have developed a Matlab package, available at \url{https://sites.google.com/site/simonbarthelme/software},
that implements EP-ABC for several models, including spatial extremes.
The current version of the package includes the parallel version described
in this paper. 

An interesting direction for future work is to integrate current developments
on model emulators into EP-ABC. Model emulators are ML algorithms
that seek to learn a tractable approximation of the likelihood surface
from samples \citep{Wilkinson:AccelABCmethodsUsingGaussianProc}.
A variant directly learns an approximation of the posterior distribution,
as in \citet{GutmannCorander:BayesOptForLikFreeInference}. \citet{Heess:LearnToPassEPMessages}
introduce a more direct way of using emulation in an EP context. Their
approach is to consider each site as a mapping between the parameters
of the pseudo-prior and the mean and covariance of the hybrid, and
to learn the parameters of that mapping. In complex but low-dimensional
models typical of ABC applications this viewpoint could be very useful
and deserves to be further explored.

\section*{Acknowledgments}

We are very grateful to Mark Beaumont, Dennis Prangle, and Markus Hainy
for their careful reading and helpful comments that helped us to improve this chapter. 

The second author is partially supported by ANR (Agence Nationale
de la Recherche) grant ANR-11-IDEX-0003/Labex Ecodec/ANR-11-LABX-0047
as part of Investissements d'Avenir program. 



\end{document}